\input harvmac.tex
\vskip 2in
\Title{\vbox{\baselineskip12pt
\hbox to \hsize{\hfill}
\hbox to \hsize{\hfill Dedicated to the memory of Ian Kogan}}}
{\vbox{\centerline{Fluid Dynamics of NSR Strings}
\vskip 0.3in
{\vbox{\centerline{}}}}}
\centerline{Dimitri Polyakov\footnote{$^\dagger$}
{dp02@aub.edu.lb}}
\medskip
\centerline{\it Center for Advanced Mathematical Sciences}
\centerline{\it and  Department of Physics }
\centerline{\it  American University of Beirut}
\centerline{\it Beirut, Lebanon}
\vskip .5in
\centerline {\bf Abstract}

We  show that the renormalization group flows
of the massless superstring modes
in the presence of fluctuating D-branes satisfy
the equations of fluid dynamics. In particular, we show
that the D-brane's U(1) field is related to the velocity
function in the Navier-Stokes equation while the
dilaton plays the role of the passive scalar advected by
the turbulent liquid. This leads us to suggest a possible
isomorphism between the off-shell superstring theory
in the presence of fluctuating D-branes and the 
fluid mechanical degrees of freedom.

\Date{June 2004}
\vfill\eject
\lref\feigen{M. Feigenbaum, J.Statist.Phys.19:25(1978)}
\lref\selff{D.Polyakov, hep-th/0401009}
\lref\krai{R. Kraichnan, Phys. Fluids 11 (1968) 945}
\lref\kraic{R. Kraichnan, Phys.Rev.Lett. 72 (1994) 1016}
\lref\gawed{K. Gawedzki, A.Kupiainen, Phys.Rev.Lett. 75 (1995) 3834}
\lref\self{D.Polyakov, Phys.Rev.D65:084041,2002, hep-th/0111227}
\lref\verlinde{H Verlinde, Phys.Lett. B192:95 (1987)}
\lref\dhoker{E D'Hoker, D.H. Phong,
hep-th/0110283, Nucl. Phys. B636:3-60 (202)}
\lref\dhokerp{E. D'Hoker, D.H. Phong, hep-th/0111016,
Nucl. Phys. B636:61 (2002)}
\lref\amps{A.M.Polyakov, Nucl.Phys.B486:23(1997)}
\lref\lpeq{A.M.Polyakov, V.Rychkov,Nucl.Phys.B581:116(2000)} 
\lref\ian{I.I. Kogan, D.Polyakov, hep-th/0212137, Phys.Atom.Nucl.66:2062
(2003)}  
\lref\kp{I.I. Kogan, D.Polyakov, hep-th/0208036, Int. J. Mod. Physics
A18:1827,2003}
\lref\fts{E. Fradkin, A.Tseytlin, Nucl. Phys B261:1-27 (1985)}
\lref\fms{D. Friedan, E. Martinec, S. Shenker, Nucl. Phys. B271:93, 1986}
\lref\ampf{S.Gubser,I.Klebanov, A.M.Polyakov,
{\bf Phys.Lett.B428:105-114}}
\lref\malda{J.Maldacena, Adv.Theor.Math.Phys.2 (1998)
231-252, hep-th/9711200}
\lref\kn{D. Knizhnik, Usp. Fiz. Nauk, 159:401-453 (1989)}
\lref\koba{Z.Koba, H.Nielsen, Nucl. Phys B12:517-536 (1969)}
\lref\witten{E.Witten, Adv. Theor. Math. Phys.2:253 (1998)}
\lref\pol{J.Polchinski, Phys.Rebv. Lett.75:4724(1995)}

\centerline{\bf Introduction}
It is common to think of strings as of objects existing
only at some exotically small scales, such as a Planck scale;
however in reality it is not necessarily so.
Perhaps the best-known example of  a non-Planckian string
is a confining (QCD) string, consisting of gluon field lines
 confined to 
narrow flux tubes connecting a pair of quarks
{\amps,\lpeq}
These thin flux tubes, which typical sizes are of the order of
$\Lambda_{QCD}$ are actually nothing but the objects
that we call the ``confining strings''.
This means that one can think of an isomorphism,
or a glossary, between string and gauge-theoretic degrees of freedom.
The purpose of this paper is to point out and explore
another example where
the stringy structures arise.Namely, we will show
how  the equations of the hydrodynamics
(e.g. Navier-Stokes and the passive scalar {\krai} equations, etc)
appear in the dynamics of strings in
fluctuating D brane-like backgrounds.
In particular, these results may imply
that one could attempt using the
language of string theory in order to study the dynamics
of vortices in a turbulent flow and , hopefully, to draw some important
conclusions about the physics of turbulence from the
the string-theoretic formalism. Clearly, the scale
at which  these ``hydrodynamical'' strings would live
 could be of the order of the lower bound
of the inertial range of the fluid dynamics.

Before we start the calculations, let us
briefly return to the QCD strings. The problem of the gauge-string
correspondence has drawn a lot of attention in the recent years and
significant progress has been  made
(especially well-known example is the AdS/CFT correspondence
observed in the supersymmetric case){\ampf,\witten,\malda}.
At this point, the string-theoretic approach remains among the most
 promising in  our attempts to understand the large N
QCD dynamics in four dimensions.

The concept of the gauge-string correspondence particularly predicts
that open QCD strings, connecting the quarks in the four-dimensional
gauge theory, actually live in the curved 5-dimensional space-time,
namely, $AdS_5$. The $AdS_5$ geometry of our space-time
is the consequence of the condition
that the partition function $Z(C)$ of an open string with its ends
attached to some contour C in four dimensions is annihilated
by the 4-dimensional loop operator, i.e. we can identify $Z(C)$
with the expectation value of the corresponding  Wilson loop
in QCD {\lpeq}.
This means that the problem of the gauge-string correspondence
is equivalent to studying the dynamics of strings in curved
e.g.  backgrounds.
Needless to say, apart from the particular problem of
the gauge-string correspondence, developing the consistent
theory of strings and D-branes in curved backgrounds
is crucial for  understanding the non-perturbative aspects
of M-theory dynamics and string dualities in general.
In general, the
 string dynamics in curved backgrounds is nonlinear and complicated and
we have little to say about the spectrum, correlation functions etc.
Our understanding of D-branes is also incomplete, 
with many results being based on arguments
coming from the low-energy effective theory.
The second quantized formalism for D-branes, in the form it is developed
for perturbative string theory, is still lacking.
Some time ago it was realized that in the spectrum of
covariant NSR superstring theory there exists a class
of physical vertex operators which  can be interpreted as second-quantized
creation operators for D-branes and used to explore
 the non-perturbative
dynamics of strings and D-branes in curved backgrounds,
while technically still working with perturbative string amplitudes
in flat space-time; in other words, these vertex operators
allow us to study strings and D-branes in curved backgrounds
by using the essentially background-independent formalism.
These vertex operators are BRST-invariant and nontrivial massless states
in NSR superstring theory,
existing at nonzero superconformal
ghost pictures only. We will refer to them as the vertices with the
 ghost-matter mixing, or the brane-like states.
Their unusual picture-dependence is due to the global singularities
 in the moduli spaces of the Riemann surfaces,
created by the insertions of these vertices on the worldsheet
{\selff}.
The picture independence of superstring amplitudes
(allowing us to move picture changing operators inside
the correlators, up to total derivatives in the moduli space)
is thus violated by these singularities.
Namely, the global singularities lead to 
the appearance  of boundaries with non-trivial topology
in the space of the supermoduli, and the total derivatives
become important, causing the picture dependence.
In the open string case, the simplest examples
of these vertices are the massles NS 5-forms,
existing at pictures $+1$ and $-3$; in the low energy limit these vertices
lead to the RR couplings of the D-branes in the DBI effective action.
The closed string brane-like states are in turn related to the kinetic
(sqrt of the determinant) terms of the DBI action.
Thus the brane-like states can be understood as D-brane creation
vertices in the second quantized formalism.
For this reason considering these vertices
as sigma-model terms in the NSR string theory is equivalent
 to introducing D-branes, or curving the background.
Thus the D-brane dynamics can be reformulated in terms of
 the scattering amplitudes involving brane-like vertex operators;
 the scattering amplitudes of the usual perturbative
string states with the brane-like insertions
will reproduce the correlators of a photon, a graviton or a dilaton
in curved backgrounds created by these D-branes.

As we shall see, the non-perturbative character
of the brane-like states is encripted 
 their non-trivial picture dependence
as well as the picture dependence of their OPE coefficients.
As the insertions of the brane-like states
are equivalent to creating D-branes and curving the backgrounds,
it is important to understand the mechanism of how the curving occurs.
For instance, the problem of interest is as follows:
suppose  that at the initial moment of time
we have a plane wave of a dilaton, propagating in
the initially flat space-time.
At  a certain moment of time a D-brane is inserted in the
background; e.g. it can be created by the relevant second quantized
vertex operator. The insertion of a D-brane leads to strong
fluctuations of the metric which later stabilizes
to some usual brane-like geometry.
The question is - how would the dilaton wave evolve in this process
of curving?
Another question of interest is the behavior of the D brane's U(1)
field in this process.
As we will see in this paper, attempting to investigate these questions
leads to the hydrodynamical equations for the RG flows
of the massless superstring modes.

In terms of the worldsheet CFT, the curving of the space-time corresponds
to the RG flow from a conformal point describing the strings in Minkowski space
to the new conformal point corresponding to the worldsheet CFT of
strings in curved brane-like geometry.
Therefore one has to study the $\beta$-functions
of the open and closed string states in the presence
of the ghost-matter mixing vertex operators.
It turns out that, because of the ghost-matter mixing,
 these $\beta$-functions are crucially different from
the usual ones in string theory;
namely  these  $\beta$-functions
 become stochastic, i.e.
contain the manifest dependence on the brane-like vertex operators,
which are the worldsheet variables. From the point of view of the
space-time fields entering the $\beta$-function equations,
the vertex operators inserted on the worldsheet are the random
variables; they can be understood as sourses of non-Markovian stochastic noise,
which memory structure is determined by the
cutoff dependence of the correlators of the ghost-matter mixing vertices.
Moreover we find that the stochastic $\beta$-function equations
for the dilaton and the photon are nothing but the equations of the
 hydrodynamics; we obtain the passive scalar equations
for the dilaton and the Navier-Stokes equation for the
stochastic vector quantity which is the D brane U(1) field
times the D-brane creating vertex operator. The incompressibility of
this ``superstringy fluid'' follows just from the transversality
of the photon.

The paper is organized as follows.
In the first section,
we review the construction of the sigma-model with the
brane-like states, starting from the integration over the
moduli and supermoduli of gravitini.
The ghost-matter mixing phenomenon and existence of the
picture-changing anomalies would follow from the singularities
in the moduli space.
In the second section, we calculate the $\beta$-functions of
the photon and the dilaton and show them to lead to
 the stochastic equations of hydrodynamics -
Navier-Stokes and the passive scalar equations accordingly.
In the concluding section we discuss possible physical
implications of our results.

\centerline{\bf NSR sigma-model and Conformal $\beta$-functions}
Consider the superstring theory
in NSR formalism, perturbed by the set $\lbrace{V_i}\rbrace$,
of massless physical vertex operators (from both the open
and closed string sectors, e.g. a photon and a dilaton)($i=1,...,n$)
The sphere or disc
scattering amplitude for N vertex operators
in the  NSR superstring theory is given by
{\verlinde, \dhoker, \selff}
:

\eqn\grav{\eqalign{
<V_1(z_1,{\bar{z_1}})...V_N(z_N,{\bar{z_N}})>
=\int{\prod_{i=1}^{M(N)}}dm_id{\bar{m_i}}\int{\prod_{a=1}^{P(N)}}d\theta_a
d{\bar{\theta_a}}\int{DX}D{\psi}D{\bar\psi}D{\lbrack{ghosts}\rbrack}
\cr
e^{-S_{NSR}+m_i<\xi^i|T_{m}+T_{gh}>+{\bar{m_i}}<{\bar{\xi^i}}|{\bar{T}}_m
+{\bar{T}}_{gh}>+
\theta_a<\chi^a|G_m+G_{gh}>+{\bar{\theta}_a}<{\bar{\chi}}_a|{\bar{G}}_m+
{\bar{G}}_{gh}>}\cr
{\prod_{a=1}^{M(N)}}\delta(<\chi^a|\beta>)\delta(<{\bar{\chi}}^a|{\bar\beta}>)
{\prod_{i=1}^{P(N)}}
<\xi^i|b><{\bar\xi}^i|{\bar{b}}>V_1(z_1,{\bar{z}}_1)...V_N(z_N,{\bar{z}}_N)}}
Here
$z_1,...,z_N$ are the points of the vertex operator insertions
on the sphere and
\eqn\grav{\eqalign{S_{NSR}\sim\int{d^2z}\lbrace
\partial{X_m}\bar\partial{X^m}+\psi_m\bar\partial\psi^m
+\bar\psi_m\partial\bar\psi^m+b\bar\partial{c}+\bar{b}\partial{\bar{c}}
+\beta\bar\partial\gamma+\bar\beta\partial\bar\gamma\rbrace\cr
m=0,...,9
}}
is the NSR superstring action in the superconformal gauge.
Next, $(m_i,\theta_a)$ are the holomorphic even and odd coordinates in the
moduli superspace and $(\xi^i,\chi^a)$ are their dual super Beltrami
differentials (similarly for $({\bar{m_i}},{\bar\theta_a})$ and
$({\bar\xi^i},{\bar\chi^a})$)
The $<...|...>$ symbol stands for the scalar
product in the Hilbert space and the delta-functions
$\delta(<\chi^a|\beta>)$ and $\delta(<\xi^i|b>)=<\xi^i|b>$
are needed to insure that the basis in the moduli space
is normal to variations along the superconformal gauge slices
(similarly for the antiholomorphic counterparts)
{\verlinde, \kn}
The dimensionalities $M(N)$ and $P(N)$
of the even and odd supermoduli spaces depend on the number
$N$ of the vertex operator insertions and are given by
{\selff}
\eqn\grav{\eqalign{
M(N)=N-3\cr
P(N)=N_{NS}+{1\over2}N_{R}+3N_{b}-2\cr
N\equiv{N_{NS}+N_{R}+N_{b}}}}
where $N_{NS}$, $N_{R}$ and $N_b$ is the
number of NS, Ramond and the brane-like vertex operator
insertions in the $N$-point amplitude.
As it has been shown in {\selff} the expression (1)
for the amplitudes leads to the following
generating functional for the $NSR$ sigma-model, involving
the set $V_i;i=1,...,n$ of the  physical vertices:
\eqn\grav{\eqalign{Z(\varphi_i)=
\int{DXD}\psi{D{\bar\psi}}{D}\lbrack{ghosts}\rbrack
{e^{-S_{NSR}+\int_k\varphi_i(k){V^i(k,z_i)}}}
\rho_{(\Gamma;Z)}\rho_{(\bar\Gamma;{\bar{Z}})}}}
where $\varphi_i$ are the space-time fields corresponding to the
 $V_i$ vertices and
$\rho_{(\Gamma,Z)}$ and its complex conjugate
$\rho_{(\bar\Gamma,\bar{Z})}$ {\self} are
 the picture-changing factors necessary to insure
the cancellations of the $\beta-\gamma$ and $b-c$ ghost
anomalies (equal to 2 for the $\beta-\gamma$ and 3 for the $b-c$
systems).
In the sigma-model (2) all the vertex operators
are to be taken in the unintegrated form, at the superconformal 
pictures $(-1)$ or ($-1,-1$) for the perturbative NS or NS-NS vertices
(such as a photon or a dilaton),
$(-1/2)$ for the Ramond  vertices and $(-3)$ for the 
brane-like vertices with the ghost-matter mixing.
The choice of the insertion coordinates $z_i$ for unintegrated vertices
is related to the choice of the  Koba-Nielsen's
measure {\koba}.
As it was shown in {\selff} such a choice of the ghost pictures
is dictated by the number of independent quadratic and $3/2$-differentials
corresponding to the vertex operator insertions,
consistent with the ghost number anomaly cancellation conditions
for the fermionic and bosonic ghosts.
The BRST-invariant
picture-changing operators for the $\beta-\gamma$ and the $b-c$-system
are given by {\fms,\selff}
\eqn\grav{\eqalign{:\Gamma:=\delta(<\chi^a|\beta>)<\chi|G>\cr
:Z:=<\xi^i|b>\delta(<\xi^i|T>)\cr
i=1...P(N);a=1...M(N)
}}
where $\chi^a$ and $\xi^i$ are the basic vectors in the 
spaces of super Beltrami differentials, dual to the odd and even
superconformal moduli.
In the delta-functional basis for $\xi^i=\delta(z-z_i)$ and $\eta^a
=\delta(z-z^a)$, the bosonic and fermionic picture-changing operators
can be expressed as
\eqn\grav{\eqalign{\Gamma(z_a)=:e^\phi(G):(z_a)\cr
Z_{open}(z_i)=\oint{{dw}\over{2i\pi}}(w-z_i)^3R(w)
Z_{closed}(z_i,{\bar{z}}_i)=\int{d^2w}|w-z_i|^6R(w)\bar{R}(\bar{w})\cr
R(w)={bT(w)-4ce^{2\chi-2\phi}TT(w)-4bc\partial{c}
e^{2\chi-2\phi}T(w)}}}
Here $G$ and $T$ are the full worldsheet matter$+$ghost supercurrent
and stress tensor, $\phi$ and $\chi$ are the bosonized ghost fields
for the $\beta-\gamma$-system.
The picture-changing operators for the $b-c$-system,
$Z_{open}$ and $Z_{closed}$  
change the $b-c$ ghost numbers of physical vertex operators
by 1 unit and 
particularly map the unintegrated
open or closed string vertices (of zero conformal dimension)  to the
vertices of the dimension 1 and $(1,1)$ respectively, integrated
over the worldsheet boundary or the entire worldshet.
It is convenient to choose $z_i$ and $z_a$ at the insertion points
of the vertex operators, corresponding to the orbifold points
in the spaces of supermoduli. In this case, the symmetry
related to the picture-changing is reduced to the discrete 
automorphism group with
 finite volume. This automorphim group includes all
the possible permutations of the picture-changing operators
between the insertion points of the vertices inside correlation functions
{\selff}.
The precise expression for the $\rho_{(\Gamma,Z)}$ in terms
of the basic vectors in the spaces of super Beltrami differentials
is given by {\selff}:
\eqn\grav{\eqalign{\rho_{(\Gamma;Z)}=
\sum_{m,n=0}^{\infty}\Xi_{\xi}^{-1}(n)\Xi_\chi^{-1}(m)
\sum_{\lbrace{\xi^{(1)},..\xi^{(n)},
\chi^{(1)}...\chi^{(m)}}\rbrace}\delta(<\chi^{(1)}|\beta>)
<\chi^{(1)}|G>...\cr
\delta(<\chi^{(m)}|\beta>)<\chi^{(m)}|G>
<\xi^{(1)}|b>\delta(<\xi^{(1)}|T>)...<\xi^{(n)}|b>\delta(<\xi^{(n)}|T>)}}
and accordingly for $\rho_{({\bar\Gamma};{\bar{Z}})}$

where the sum over $\xi^{(i)}$ and $\chi^{(i)}$ implies the summation over all
the basic vectors of the $(m,n)$-dimensional
spaces of the differentials;
$\Xi_\chi$ and $\Xi_\xi$ are the volumes of the automorphism groups
corresponding to the permutations of 
$\Gamma$ and $Z$ picture-changing operators
between the vertices.
Factorization by these volumes
For example, for the scattering
amplitude of N gravitons on the sphere one
has
$\Xi_\xi=N^{N-3},\Xi_\chi=N^{N-2}$
Structurally, it is easy to see from (7) that the $\rho$-factor
has the form:
 the $\rho_{(\Gamma,Z)}\sim({1+:\Gamma:+:\Gamma^2:+:\Gamma^3:+...})
(1+:Z:+:Z^2:+:Z^3:+...)$ 

As the pictures of the vertex operators in the sigma-model (4)
are fixed from the beginning, the role of the $\rho$-factor is to insure
 the ghost number anomaly cancellation in the amplitudes.
For example, in case of the scattering of N unintegrated gravitons 
on the sphere the only term in the $\rho_{(\Gamma,Z)}$-expansion
contributing to the correlator is of the structure
${\sim}:\Gamma^{N-2}Z^{N-3}:$, while contributions
from other terms are absent due to the ghost number balance.
In other words, each N-point amplitude  automatically picks up the 
appropriate terms from the $\rho$-factor, 
to insure the correct overall ghost number
and the factorization condition.
Dividing by the number of the permutations of the p.c. 
operators between the vertices
(equal to the volume of the automorphism group for a particular amplitude)
leads to the appropriate normalization of the correlator 
and the corresponding terms in the low-energy effective action.
Thus in case of the usual perturbative vertices the 
$\rho_{(\Gamma,Z)}$-factor simply insures the correct normalization and 
ghost anomaly anomaly cancellation in the amplitudes and the
conformal $\beta$-functions
; in  the picture-dependent case of the brane-like vertices
(corresponding to global singularities of the moduli spaces)
things become more subtle and complicated.In the next section we will 
demonstrate how the picture-dependence of the OPE coefficients leads
to the stochastic terms in the stochastic RG equations

\centerline{\bf Picture-dependent OPE and Stochastic $\beta$-functions}

Consider the expansion of the generating functional (4) of the sigma-model
 in terms of $\varphi_i$ space-time fields.We get
\eqn\grav{\eqalign{Z=\int{D\lbrack{X},\psi,{\bar\psi},ghosts\rbrack}
e^{-S_{NSR}}\rho_{(\Gamma,Z)}\rho_{(\bar\Gamma,\bar{Z})}
\cr(1+\varphi_i{V_i}
+{1\over2}\varphi_i\varphi_j{V_iV_j}+{1\over6}\varphi_i\varphi_j\varphi_k
V_iV_jV_k+...\cr
i,j,k=1,...n)}}
The Z-transformation of the vertices (by the appropriate
expansion terms from $\rho_{\Gamma,Z}$ and its complex conjugate)
leads to {\selff}
\eqn\grav{\eqalign{:Z{\bar{Z}}V_i:\sim\int{d^2z}W_i(z,\bar{z})
+\lbrace{Q_{brst},...}\rbrace\cr
i=1,...n}}
for the closed string vertices and
\eqn\grav{\eqalign{:Z{\bar{Z}}V_i:\sim\int{{d\tau}\over{2i\pi}}W_i(\tau)
+\lbrace{Q_{brst},...}\rbrace\cr
i=1,...n}}
for open strings.
where $W_i$ are the dimension 1 or (1,1) operators, 
which worldsheet integration gives the integrated vertices for
open and closed strings.
Now consider the quadratic term in the expansion (8)
which, upon the Fourier transform, is proportional to 
$\sim\int_{p,q}
{\varphi_i(p)\varphi_j(q)\int{d^2z}\int{d_2w}W_i(z,\bar{z};p)W_j(w,\bar{w};q)}$
where $\int_{p,q}$  stands for the integral over the momenta
$p$ and $q$ of the vertices
(the open string case can be treated analogously).
In general, the OPE between $W_i$ and $W_j$ is singular and is given by.
\eqn\lowen{W_i(z,\bar{z};k)W_j(w,\bar{w};p)\sim{1\over{|z-w|^2}}
C_{ij}^k(p;q)W_k({{z+w}\over2},{{{\bar{z}}+{\bar{w}}}\over2};p+q)+...}
where $C_{ij}^k$ are the structure constants and the summation 
over $k$ is from 1 to n (i.e. over all the physical massless
vertex operators). This OPE singularity leads to the divergence
in the partition function as
\eqn\grav{\eqalign{\int_{p,q}
\int{d^2z}\int{d^2w}W_i(z,\bar{z};p)W_j(w,\bar{w};q)
\cr
\sim\int_{k_1,k_2}{C_{ij}^k}(k_1,k_2)
\int{{d^2\eta}\over{|\eta|^2}}\int{d^2\xi}W_k(\xi,{\bar\xi};k_1)
\cr
\equiv{log{\Lambda}}
\int_{k_1,k_2}{C_{ij}^k}({{k_1+k_2}\over2},{{k_1-k_2}\over2})
\int{d^2\xi}W_k(\xi,{\bar\xi};k_1)}}
where $\Lambda$ is the worldsheet cutoff and
\eqn\grav{\eqalign{k_1=p+q\cr
k_2=p-q\cr
\xi={{z+w}\over2}\cr
\eta={{z-w}\over2}}}
To eliminate this divergence one has to renormalize the
space-time field $\varphi_k$ in the linear term in the expansion (8)
as
\eqn\lowen{\varphi^k(p)\rightarrow\varphi^k(p)-
{1\over2}log{\Lambda}\int_{q}C_{ij}^k({{p+q}\over2},{{p-q}\over2})
\varphi_i({{p+q}\over2})\varphi_j({{p-q}\over2})}
The conformal invariance condition on the worldsheet 
(i.e. the cutoff independence of the partition function)
implies the low-energy effective equations of motion for the  
space-time fields:
\eqn\lowen{{\int_q}C_{ij}^k({{p+q}\over2},{{p-q}\over2})
\varphi_i({{p+q}\over2})\varphi_j({{p-q}\over2})=0}
This particularly implies the Einstein's equations
for the closed string massless fields and the
 DBI equations of motion for the photon {\fts}.
Next, the renormalization group flow (14) generated by the OPE singularity
(12) also deforms the next order,quadratic term in the expansion (8).
Under the renormalization (14) the quadratic term flows 
into the divergence cubic in $\varphi$ as
\eqn\grav{\eqalign{{1\over2}\int_{k,p}\varphi_i(k)\varphi_j(p)
\int{d^2z}\int{d^2w}W_i(z,{\bar{z}})W_j(w,{\bar{w}})
\cr\rightarrow-{1\over2}log{\Lambda}
\int_{k,p,q}
C_{ij}^k({{p+q}\over2},{{p-q}\over2})
\varphi_i(k)\varphi_j({{p+q}\over2})\varphi_k({{p-q}\over2})
\cr\times
\int{d^2z}\int{d^2w}W_i(z,{\bar{z}};k)W_j(w,{\bar{w}};p)}}
This divergent term, however, is precisely cancelled
by the divergency stemming from the OPE singularities
inside the cubic term of the expansion (8). Indeed, using
the OPE (11) it is easy to show that
\eqn\grav{\eqalign{
{1\over6}\int_{k,p,q}
\varphi_{i}(k)\varphi_j(p)\varphi_k(q)\int{d^2z}
\int{d^2w}\int{d^2u}V_i(z,{\bar{z}};k)V_j(w,{\bar{w}};p)
V_k(u,{\bar{u}};q)\cr\rightarrow
{1\over2}log{\Lambda}
\int_{k,p,q}
C_{ij}^k({{p+q}\over2},{{p-q}\over2})
\varphi_i(k)\varphi_j({{p+q}\over2})\varphi_k({{p-q}\over2})
\cr\times
\int{d^2z}\int{d^2w}W_i(z,{\bar{z}};k)W_j(w,{\bar{w}};p)}}
The cancellation of the divergences (16) and (17), stemming from the flow
of the quadratic term and the OPE singularities of cubic term, 
is important as it insures that the RG equations for the space-time
fields $\varphi_i$ are deterministic. Indeed, the presence of
 the divergencies of the type $\sim{log}\Lambda\varphi_i\varphi_j
\varphi_k{C_{ij}}^k\int{W_iW_j}$ in the expansion (8) would lead
to appearance of the extra terms  in conformal $\beta$-functions for the
space-time fields ${{d\varphi_i}\over{d{log\Lambda}}}$, proportional
to $\sim{C_{jk}^l}\varphi_i\varphi_j\varphi_k\int_\Lambda{V_l}$
with the worldsheet integral of the  $V_l$ vertex operators
cut off at the scale $\Lambda$.
In other words, the RG equations for the space-time 
fields would manifestly depend on the worldsheet variables!
From the point of view of the cutoff-dependent space-time
physics, the vertex operators,involving the random functions
defined on the worldsheet
(such as the coordinate $X(z,{\bar{z}})$) play the role of the 
non-Markovian stochastic noise. 
Indeed, the $\beta$-function equations
for the space-time fields would have the form of the Langevin stochastic 
equation:
\eqn\lowen{{{d\varphi}\over{d\tau}}={{\delta{S}(\varphi)}\over{d\varphi}}
+{C}\varphi^3\eta(\tau)}
where
$C$ are the structure constants,
the stochastic time variable $\tau\equiv{log{\Lambda}}$
is defined by the logarithm of the worldsheet cutoff and
 the role of the stochastic noise
$\eta\equiv\int_\Lambda{d^2z}{V(z,{\bar{z}})}$ 
is played by the worldsheet integral of $V$, cut off at a scale $\Lambda$.
The memory structure  of the noise is defined by the worldsheet correlations
of the $V$ operators.
In any case, as we have already noted above, the stochastic terms
of the form (18) do not appear in the $\beta$-function equation
in the standard string perturbation theory as the divergencies
 (16) and (17) cancel each other.
At the first glance, the cancellation of (16) and (17),
insuring the determinism of the worldsheet $\beta$-functions,
is quite trivial and automatic, true for all the
 orders of the perturbative expansion.
 Things, however, become far more subtle for the cases when 
the structure constants are picture-dependent, i.e.
the OPE $V_iV_j\sim{C}V_k$ depends on the  ghost pictures
at which $V_i$ and $V_j$ are taken.
For usual perturbative vertices, such as a photon or a graviton,
such a picture dependence is of course absent and accordingly
their $\beta$-functions are deterministic.
The situation is drastically different in the case
of the ghost-matter mixing, i.e. in the presence of the
vertex operators breaking the equivalence of the ghost pictures 
{\selff, \self}

The space-time fields corresponding to these vertices usually correspond to
 the collective coordinates of D-branes - their effective actions
are shown to be of the DBI type; one can also show
that these operators carry nonzero RR charges by computing their correlators
with the RR-vertices {\kp}.
In fact,
 these vertex operators can be thought of as the D-branes in the 
second-quantized formalism {\kp}.
Important examples of such operators (at the integrated
$b-c$ picture)
are given by:
\eqn\grav{\eqalign{V_5^{o.s.(-3)}=R_{m_1...m_5}(k)\oint{{d\tau}\over{2i\pi}}
e^{-3\phi}\psi_{m_1}...\psi_{m_5}e^{ikX}\cr
V_5^{o.s.(+1)}=R_{m_1...m_5}(k)\oint{{d\tau}\over{2i\pi}}
e^{\phi}\psi_{m_1}...\psi_{m_5}e^{ikX}+ghosts}}
for open strings
and
\eqn\grav{\eqalign{{V_5}{\equiv}V_5^{{c.s}{(-3)}}
=H_{m_1...m_6}(k)\int{d^2z}e^{-3\phi-\bar\phi}\psi_{m_1}
...\psi_{m_5}\bar\psi_{m_6}{e^{ikX}}(z,{\bar{z}})\cr
V_5^{{c.s}(+1)}=H_{m_1...m_6}(k)\int{d^2z}e^{\phi+\bar\phi}
\psi_{m_1}...\psi_{m_5}\bar\psi_{m_6}e^{ikX}(z,{\bar{z}})
+ghosts}}
Here and elsewhere the upper indices in the round  brackets label the 
$\beta-\gamma$ ghost pictures.
The open string $V_5$-operators 
exist at pictures $-3$ and below as well as $+1$ and above, but not at the
pictures $-2$, $-1$ and $0$.
In cases when the $V_5$-operators are taken at the positive pictures
($+1$ and above), they must include the $b-c$ ghost extra terms,
to insure their BRST-invariance (apart from the $e^\phi{\psi}^5$-part)
(see {\self} or {\selff} for the precise expressions
for these terms). All the higher positive pictures
for $V_5$-operators can be obtained by applying the 
standard picture-changing to 
$V_5^{(+1)}$ while the lower negative pictures
for $V_5$ can be derived using
the inverse picture-changing of $V_5^{(-3)}$.
$V_5^{(+1)}$ can be obtained from $V_5^{(-3)}$
by making the Z-transformation first
(i.e. putting $V_5^{(-3)}$ in the ``double-integrated'' form)
and the subsequent application of the fourth power of the usual
picture-changing operator.
The closed-string $V_5$-operators (20) can 
be obtained straightforwardly from the open string ones
by multiplying them by the antiholomorphic photonic part
(which of course can be taken at any picture).
The BRST invariance and non-triviality
conditions for the operators (19),(20) are given by {\kp, \selff}:
\eqn\grav{\eqalign{k_{{\lbrack}m_6}R_{m_1...m_5\rbrack}(k)\neq{0}\cr
k_{{\lbrack}m_7}H_{m_1...m_5{\rbrack}m_6}(k){\neq}0\cr
k_{m_6}H_{m_1...m_5m_6}(k)=0}}
where the $H$-tensor is antisymmetric in the first 5 indices and the
square brackets imply the antisymmetrization.
In particular,
the BRST conditions (21) can be shown
to reduce the number of independent components of the $H$-tensor by one half,
i.e. the number of the physical d.o.f. related to the $H$-tensor is equal to
 1260. In addition, these BRST conditions imply that
for each particular polarization $m_1...m_6$ of the $H$-tensor
the momentum $k$ must be orthogonal to the $m_1,...,m_6$ directions in the 
space-time.
As the number of independent polarizations
is given by ${{10!}\over{4!6!}}=210$, the number of the physical d.o.f.
per polarization is equal to 6.
By computing the low-energy effective action for the $V_5$-states
with a fixed polarization  one can show that these degrees of freedom
correspond to 6 collective coordinates for a D3 brane's transverse fluctuations
which can be parametrized as
\eqn\brav{\eqalign{
\lambda_t\equiv{H_{t_1...t_5t}}\cr
t=4,...,9;t{\neq}t_1,...,t_5}}
(H is antisymmetric in $t_1,...,t_5$)
and the low-energy limit is described by the
DBI action for D3-brane with the worldvolume in 
the $0,...,3$ directions, in terms of $\lambda_t$ {\selff,\kp}
In the spherically symmetric case (
s-wave approximation in the near-horizon limit)
one can take all the components equal
\eqn\lowen{\lambda_t(k)\equiv\lambda(k);t=4,...9}
To simplify things, in the rest of this paper, the $V_5$-excitation,
entering the NSR sigma-model, will be taken in the form (23).
As the OPE's involving these operators are picture-dependent
(see the discussion below) one must accurately  account for
 the $\beta$-function contributions 
from various ghost pictures in the expansion (8).
Then, as a result of the picture asymmetry in the operator products,
the renormalization of the cubic terms won't necessarily 
cancel  the flow of the quadratic ones, as in 
the case of (16) and (17).
As a result, the resulting $\beta$-function equations would 
be operator-valued, or stochastic.
In particular, in the absence of the U(1) gauge field, 
this stochastic process leads to the evolution of the space-time geometry
from flat to the one with the D-brane metric, corresponding to the appropriate
DBI effective action (see the discussion above). In other words,
the D-brane creation operators are inserted in the the theory
with the background originally flat; 
such an insertion leads to the non-Markovian
stochastic process which thermodynamical limit is given by the
string theory in the curved D-brane type geometry.
As a result, exploring the correlation function involving
the brane-like states allows us to study the dynamics of 
strings in curved backgrounds, while technically working with
string perturbation theory arond the flat vacuum.
 Such an approach is referred to as the ghost-matter mixing formalism
{\self,\kp}
In the presence of the U(1) gauge field, the stochastic 
$\beta$-function equations,involving the brane-like states, become more 
complicated and take the form of the equations of hydrodynamics.
The rest of the paper will discuss the derivation of these equation
from the sigma-model for NSR strings, as well as the 
related physical implications.

\centerline{\bf $\beta$-functions in the D-brane Backgrounds
and Equations of Hydrodynamics} 
Consider the NSR superstring theory in the
background of standard perturbative states
(a photon, a dilaton,a graviton and an axion)
 and the closed string
$V_5$-excitation (20),(22), (23) corresponding to the D3-brane insertion.
For simplicity, let the polarization and the propagation of the
perturbative states be confined to the four-dimensional subspace,
defined by the admissible directions of the momentum of the
$V_5$-vertex, following from the BRST conditions (21)
(corresponding to the orientation of the D3-brane's worldvolume) 
 
The generating functional of this model is given by
\eqn\grav{\eqalign{Z(\lambda,\varphi,A_a,H_{ab})=
\int{D}{\lbrack}X,\Psi,\bar\Psi, ghosts\rbrack
exp\lbrace
{-S_{NSR}}
+\int_k{\lbrace}A_a(k)V_{ph}^a(\tau;k)
\cr
+H^{ab}{V_{ab}}
(z,{\bar{z}};k)
+{\int_p}\lambda(p)V_5^{(-3)}(w,{\bar{w}};p){\rbrace}|
\rho_{(\Gamma,Z)}|^2
}}
with
\eqn\grav{\eqalign{V_{ph}(\tau)=ce^{-\phi}\psi^a{e^{ikX}}(\tau)\cr
V^{ab}(z,{\bar{z}};k)=c{\bar{c}}
e^{-\varphi-\bar\varphi}\psi^a\bar\psi^b
{e^{ikX}}(z,{\bar{z}})\cr
V_5^{(-3)}(w,{\bar{w}};p)=c{\bar{c}}e^{-3\phi-\bar\phi}
\psi_{\lbrack{4}}...\psi_8{\bar\psi}_{9\rbrack}e^{ip_a{X^a}}
(w,{\bar{w}})\cr
a=0,...,3
}}
and the rank 2 massless field $H_{ab}$ being either a dilaton, a graviton
or an axion:
\eqn\grav{\eqalign{
H_{ab}=G_{ab}(k)+B_{ab}(k)+\varphi(k)(\eta_{ab}-k_a{\bar{k}}_b
-{\bar{k}}_a{k_b})\cr
k^2={\bar{k}}^2=0;(k{\bar{k}})=1}}
Consider the expansion of this functional in the space-time fields.
Namely, consider the term quadratic
in $\lambda$ and $A_m$, and of the arbitrary order $N$ in the
closed string $H$-field.
This term is given by the correlation function:
\eqn\grav{\eqalign{A(p_1,p_2;k_1,k_2;q_1,...,q_N)\cr\sim
{1\over{(2!)^2N!}}<|\rho_{(\Gamma;Z)}|^2
V_5^{(-3)}(p_1)V_5^{(-3)}(p_2)V^a(k_1)V^b(k_2)
V^{a_1b_1}(q_1)...V^{a_Nb_N}(q_N)>}}
(all the operators except for the $V_5$-vertices are
at the left picture -1 and all the right pictures here
and elsewhere are assumed to be $-1$ unless specified otherwise) 

We need to find the contribution
of this correlator to the photon's beta function
in the presence of the $V_5$-insertions (i.e. in the D3-brane background)
This contribution is determined by  
the OPE singularities between the $V_5$- operators and the photon 
vertices inside the correlator (27).
Clearly, such a contribution must depend only on the
structure constants  of this operator product  and not on any other 
insertions not involved in the OPE (in particular, 
the $N$-independence is the 
consequence of the factorization conditions for the 
string amplitudes, as well as the condition for the renormalizability
of the string perturbation theory).
In case when the structure-constants are picture-independent,
such an $N$-independence is quite obvious - this is why in the 
$\beta$-function calculations (14), (16), (17)
  it is sufficient to consider the OPE of 
a couple of the vertex operators $V_i$ and $V_j$ regardless 
of details of particular correlators they belong to -
 normalization and unitarity conditions would always insure that
the RG flow (14) of the space-time fields would simultaneously cancel
the divergencies stemming from the OPE (11) inside all the correlators.
In the picture-dependent case involving the $V_5$-operators the 
situation is far more subtle and the perturbation theory must be modified
in order to satisfy the conditions of the renormalizability and
the factorization of the amplitudes. Let us now concentrate
on the scattering amplitude (27)
The relevant term from $\rho_{(\Gamma,Z)}\rho_{(\bar\Gamma\bar{Z})}$
to cancel the left and right ghost number anomalies
has the structure
 $\sim:(Z{\bar{Z}})^{N+1}::\Gamma^{N+6}:{:{\bar{\Gamma}}:}^{N+2}$.
The volumes of the automorphism groups related to the left and
right $\Gamma$ and $Z$-permutations between the insertion points
of the $N+4$-point correlator (27) with two $V_5$-vertices
are computed to be
\eqn\grav{\eqalign{
\Xi_\Gamma(N+4;2)=(N+4)^{N+6}\cr
-2(N+3)^{N+6}{\lbrace}
{{N+6}\over{N+3}}+{{(N+5)(N+6)}\over{2(N+3)^2}}
+{{(N+4)(N+5)(N+6)}\over{6(N+3)^3}}\rbrace\cr
+(N+2)^{N+7}{\lbrace}
{{(N+5)(N+6)}\over{(N+2)^2}}+{{(N+4)(N+5)(N+6)}\over{(N+3)^3}}
\cr
+{{7}\over{12}}{{(N+3)(N+4)(N+5)(N+6)}\over{(N+2)^4}}
+{1\over{36}}{{(N+1)(N+3)(N+4)(N+5)(N+6)}\over{(N+2)^5}}\rbrace\cr
\Xi_{{\bar{\Gamma}}}(N+4;2)=(N+4)^{N+2}\cr
\Xi_\xi(N+4;2)=\Xi_{\bar{\xi}}(N+4;2)=(N+4)^{N+1}}}
Factorizing by these volumes in the $|\rho_{(\Gamma,Z)}|^2$-factor
insures the correct normalization of the
related terms in the low-energy effective action, as well as the
amplitude factorization and unitarity conditions
( after summing over all  
the admissible picture configurations
for the vertices in the  amplitude (27), equivalent 
to the admissible choices of basis in the supermoduli space
associated with this amplitude)
If the $V_5$-operators were picture-independent, the $\Xi$-volumes 
for $N+4$-point amplitude would have been given simply by
the numbers of the permutations  of 
the $N+1$ picture-changing Z-operators and $N+2$ $\Gamma$-operators
between  $N+4$ insertion points, i.e. $(N+4)^{N+1}$
and $(N+4)^{N+2}$ respectively.
However, because of the  $\beta-\gamma$ picture-dependence of
the  $V_5$-insertions in the $N+4$-point function,
while the $\Xi_\xi$-volume is still unchanged,
the $\Xi_\Gamma$ volume related to the left $\Gamma$-permutations 
must exclude the combinations leading to the non-existent
$0$,$-1$ and $-2$-pictures for any of the $V_5$-insertions
(recall that the $V_5$-operators are originally taken at the picture
$-3,-1$ while the rest are at the picture $-1,-1$).
 A simple calculation then leads to (28), insuring that
the  factorization and unitarity conditions for the amplitude (27)
are satisfied.
Next, consider the OPEs between the $V_5$ and $V^a$
contributing to the photon's $\beta$-function.
As has been noted above, these OPE's are picture-dependent.
For example, the OPE of the picture 0  photon with the $V_5$-operator gives
\eqn\grav{\eqalign{lim_{z,{\bar{z}}\rightarrow{\sigma}}
V^{{a}{(0)}}(k,\tau)V_5^{(-3,-1)}(p,z,{\bar{z}})
\cr
=
(\partial{X^a}+i(k\phi)\psi^a)e^{ikX}(\tau)e^{-3\phi-\bar\phi}
(\psi_{t_1}...\psi_{t_5}\bar\psi_{t_6}+c.c.)e^{ipX}(z,\bar{z})\cr
\sim{ip^a}({1\over{\tau-z}}+{1\over{\tau-{\bar{z}}}})V_5(k+p)
=2ip^a(\tau-\sigma)^{-1}V_5(k+p)}}
and contributes to the photon's $\beta$-function since it is singular 
as $z,{\bar{z}}\rightarrow{\sigma}$
where $\sigma$ is some point on the worldsheet boundary, close to $\tau$.
At the same time, the 
integrand of the photon's vertex operator at the picture $+1$
is given by
\eqn\grav{\eqalign{
V^{{a}{(+1)}}(k,\tau)=-{1\over2}
e^{\phi}e^{ikX}\lbrace{((\psi\partial{X})(\partial{X^a})
+i(k\psi)\psi^a)}+{1\over2}\partial^2\psi^a+\partial\psi^a{P^{(1)}_{\phi-\chi}}
\cr
+\psi^a{P^{(2)}_{\phi-\chi}}+i(k\psi)\partial{X^a}P^{(1)}_{\phi-\chi}
+i(k\partial\psi)\partial{X^a}\cr
+i(k\partial{X})P^{(1)}_{\phi-\chi}
\psi^a+i(k\partial^2{X})\psi^a+i(k\partial\psi)(k\psi)\psi^a
\cr
+i(k\psi)(\partial^2{X^a}+P_{\phi-\chi}^{(1)}\partial{X^a})\rbrace
-{1\over4}be^{2\phi-\chi}e^{ikX}
P^{(2)}_{2\phi-2\chi-\sigma}(\partial{X^a}
+i(k\psi)\psi^a)}}
where we have defined
\eqn\lowen{P_{f}^{(n)}={1\over{n!}}e^{-f}{{d^n}\over{dx^n}}e^{f}}
for any $f(x)$. It is now easy to check that the OPE of the
picture $+1$ photon with $V_5$ is non-singular:
\eqn\lowen{{lim_{z,{\bar{z}}\rightarrow{\sigma}}}V^{a(+1)}(k,\tau)V_5^{(-3,-1)}(z,\bar{z})\sim
O(|\tau-\sigma|^0)}
and therefore does not contribute to the $\beta$-function.
In the general case, one can show that the picture-dependent 
OPE between the photon and $V_5$ is given by 
\eqn\grav{\eqalign{
V_5^{(s_1)}(z,{\bar{z}};p)V^{a(s_2)}(\tau;k)\cr\sim
i({1\over{z-\tau}}+{1\over{\bar{z}-\tau}})p^a{C^{(s_1|s_1+s_2)}}
V_5^{(s_1+s_2)}({{z+\tau}\over2},{{{\bar{z}}+\tau}\over2};k+p)
+...\cr
V_5^{(s_1)}(z,{\bar{z}};k)V_5^{(s_2)}(w,{\bar{w}};p)
\sim{{C^{(s_1|s_2)}}\over{|z-w|^2}}(
(kp)V_\varphi^{(s_1+s_2)}({{z_1+z_2}\over2},{{{\bar{z}}_1+{\bar{z}}_2}\over2};
k+p)\cr+
{{2k_m}\over{z-{\bar{z}}+w-{\bar{w}}}}V^{m(s_1,s_2)}(\tau;k+p))}}
where we have skipped the non-singular part of the OPE.
The $s_i$ indices label the pictures of the vertex operators
and the symmetric picture matrix $C^{(s|t)}$ reflects the picture dependence
of the OPE coefficients in the singular terms. Namely,
$C^{(s|t)}=0$ if either of the pictures $s,t=0,-1,-2$
and 1 otherwise.
Let us now return to the amplitude (27).
Summing over the admissible configurations of the basic vectors 
in the supermoduli space, stemming
the $\rho_{(\Gamma,Z)}$-factor, we obtain
\eqn\grav{\eqalign{A(p_1,p_2;k_1,k_2;q_1,...,q_N)\sim{1\over{(2!)^2N!}}
\Xi_\Gamma^{-1}(N+4;2)\sum_{s_1,s_2,t_1,t_2,r_1,...r_{n-1};s_{1,2}{\neq}
1,2,3}^{s_1+s_2+t_1+t_2+\sum_{j=1}^N{r_j}=N+6}\cr
{{(N+6)!}\over{s_1!s_2!t_1!t_2!r_1!...r_{N-1}!(N+6-s_1-s_2-t_1-t_2-\sum{r})!}
}\cr
<V_5^{(s_1-3)}(p_1)V_5^{(s_2-3)}(p_2)V_a^{(t_1-1)}(k_1)
V_b^{(t_2-2)}(k_2)
{\prod_{j=1}^{N}}V_{a_jb_j}^{(r_j-1)}(q_j)>}}
where we suppressed the antiholomorphic picture indices
for the closed string operators (note that the antiholomorphic
part of the amplitude is picture-independent and the effect
of the $\rho_{({\bar\Gamma},{\bar{Z}})}$-factor is trivial;
summing over pictures and dividing by the
automorphism volume related to the antiholomorphic pictures, results 
 in the trivial
 unit normalization factor  for the right-moving part of the amplitude,
as in any picture-independent case).
The sum (34) gives the total contribution to the scattering amplitude 
(27) from all admissible picture configurations.

To evaluate the sum (34), note that, even though
 the operator algebra involving the ghost-matter mixing vertices is 
generally picture-dependent (see the discussion below), all  
the $N+4$-point correlators entering the sum (34) have the same value
for
$s_1+t_1{\neq}0,{-1,-2};s_2{\neq}0,-1,-2$:
\eqn\grav{\eqalign{<V_5^{(s_1-3)}(p_1)V_5^{(s_2-3)}(p_2)V_a^{(t_1-1)}(k_1)
V_b^{(t_2-1)}(k_2)
V_{a_1b_1}^{(r_1-1)}(q_1)...V_{a_Nb_N}^{(r_N-1)}(q_N)>
\cr\equiv{S_{N+4}}(k_1,k_2,p_1,p_2,q_1...q_N)}}
This is because the full operator algebra for two $V_5$-operators
has the form
\eqn\grav{\eqalign{V_5^{(s)}V_5^{(t)}\sim\sum_{X,Y}C_X\lbrack{X^{(s+t)}}
\rbrack
+C_Y^{(s|t)}\lbrack{Y^{(s+t)}}\rbrack\cr
V_5^{(s)}\lbrack{X^{(t)}}\rbrack\sim\sum_{Y}C_5^{(s|t)}\lbrack{Y^{(s+t)}}
\rbrack\cr
s,t\neq{-2,-1,0}
}}
where $\lbrack{X}\rbrack$ is the subclass of the picture-independent operators
(corresponding to the tower of standard perturbative
massless and massive superstring states)
and $\lbrack{Y}\rbrack$ is the tower of picture-dependent
(brane-like) vertices describing the non-perturbative
sector of superstring fluctuations.

Note that the $C_Y$ and $C_5$ coefficients are picture dependent
while $C_X$ are not. 
Then the second OPE of (36) insures that the
$\lbrack{Y}\rbrack$-part of the first OPE (36) with the 
picture-dependent $C_Y$ coefficients does not contribute to the correlators
with two $V_5$-insertions,
and the picture independence of the $C_X$ coefficients
for $s,t{\neq}-2,-1,0$
insures the independence of the scattering amplitude (35)
on the choice of $s,t,r$-pictures, 
insuring its unitarity and factorization properties.
Note that the operator product expansion of any operators
from the $\lbrack{X}\rbrack$-subclass does not contain
any operators from the $\lbrack{Y}\rbrack$-subclass
(but the reverse is not true). This is why the brane-like
states never appear as the intermediate poles in the  perturbative
amplitudes and do not affect the standard string perturbation theory.
Using (35) we can now evaluate the sum (34) obtaining
\eqn\lowen{A(p_1,p_2,k_1,k_2,q_1,...q_N)={{1\over{{(2)!}^2N!}}}
S_{N+4}({p_1,p_2,k_1,k_2,q_1,...q_N})}
i.e. the presence of the $\Xi_\Gamma^{-1}(N+4;2)$-factor
in the expression (34) insures the correct normalization of the 
scattering amplitude and the corresponding term in the low energy .
in the low-energy effective action.
 Next, substituting the OPE (33) into the amplitude (34)
(multiplying the latter by the corresponding space-time fields)  
and performing the worldsheet 
integration we find that the OPE singularity leads to the 
logarithmic divergence
in this correlator (and hence in the partition function), given by
\eqn\grav{\eqalign{A_\Lambda(p_1,p_2;k_1,k_2;q_1...q_N)=(ip_{1a})log\Lambda
{{\lambda^2{A_a}A_bH_{a_1b_1}...H_{a_Nb_N}}\over{(2!)^2N!\Xi_\Gamma(N+2;4)}}
\cr\sum_{s,t,r}
{{(N+6)!}\over{s_1!s_2!t_1!t_2!r_1!...r_{N-1}!(N+6-s_1-s_2-t_1-t_2-\sum{r})!}
}C^{(s_1-3|s_1+t_1-4)}\cr
<V_5^{(s_1+t_1)}(p_1+k_1)V_5^{(s_2)}(p_2)V_b^{(t_2-1)}(k_2)
V_{a_1b_1}^{(r_1-1)}(q_1)...>\cr+
perm.{\lbrace}(s_1\leftrightarrow{s_2};p_1\leftrightarrow{p_2})
+(t_1\leftrightarrow{t_2};k_1\leftrightarrow{k_2})+
(s_1,t_1\leftrightarrow{s_2,t_2};k_1,p_1\leftrightarrow{k_2,p_2})}}
Again, using (35) and substituting for the $C^{(s|t)}$ picture matrix
we can perform the summation in (38) leading to
\eqn\grav{\eqalign{A_\Lambda(p_1,p_2;k_1,k_2;q_1...q_N)=
(ip_{1a})log\Lambda{{\lambda^2A_aA_bH_{a_1b_1}...H_{a_Nb_N}}\over
{2!N!}}\cr{\times}
(1-\Xi^{-1}_\Gamma(N+4;2)F(N))S_{N+3}(k_1+p_1,p_2,k_2,q_1...q_N)}}
where
\eqn\grav{\eqalign{S_{N+3}(p_1,p_2,k,q_1...q_N)\cr
\equiv
<V_5^{(s_1-3)}(p_1)
V_5^{(s_2-3)}(p_2)V_b^{(t-1)}(k)V_{a_1b_1}^{(r_1-1)}(q_1)...
V_{a_Nb_N}^{(r_N-1)}(q_N)>
\cr
s_{1,2},t,r_1,...r_{N}\geq{0}
\cr
s_{1,2}\neq{1,2,3};\cr
s_1+s_2+t+{\sum_{j=1}^N}r_j=N+5
}}
$\Xi_\Gamma$ is defined by (28)
and the function $F(N)$ is given by
\eqn\grav{\eqalign{
F(N)=
(N+2)^{N+6}{\lbrace}
{1\over2}{{(N+5)(N+6)}\over{(N+2)^2}}+
{1\over6}{{(N+4)(N+5)(N+6)}\over{(N+2)^3}}
(1+{1\over2}{{N+3}\over{N+2}})\cr
-{{(N+3)(N+4)(N+5)(N+6)}\over{(N+2)^4}}
({{17}\over{72}}({{N+1}\over{N+2}})^{N+1}+
{1\over2}({{N+1}\over{N+2}})^{N+2}
)\cr
-{1\over{72}}
{{N(N+3)(N+4)(N+5)(N+6)}\over{(N+2)^5}}({{N+1}\over{N+2}})^{N}
\cr
-{1\over2}{{(N+4)(N+5)(N+6)}\over{(N+2)^3}}({{N+1}\over{N+2}})^{N+3}
\rbrace
}}
As a result, the picture dependence of the OPE (33)
leads to the dependence of the normalization of
logarithmic divergence (38)
on the number of external insertions $V_{a_jb_j}$
(the operators other than the $V_5$ and the photons)
Obviously, in the absence of the picture dependence
(implying $\Xi_\Gamma(N+2;4)=(N+4)^{N+6};C^{(s|t)}{\equiv}1$ for any $s,t$
and $F(N)\equiv{0}$)
there would be no dependence on $N$ and
this divergence could have been removed by the renormalization of the
$\lambda$-field in the lower order
$N+3$-point correlation function
 proportional to
\eqn\lowen{S_{N+3}(p_1,p_2;k_1;q_1,...,q_N)\sim{A}\lambda^2H^{n}
{<V_5(p_1)V_5(p_2)V_a(k_1)V_{a_1b_1}(q_1)...V_{a_Nb_N}
(q_N)>}} The naive renormalization
of the $\lambda$-field removing the divergence would have been
schematically given by
(suppressing the momentum integration) 
\eqn\lowen{\lambda\rightarrow{\lambda}-C_aA_a\lambda{log}\Lambda}
($C_a$ are the structure constants), 
leading to the standard terms in the $\beta$-function of the 
 $\lambda$ field. 
In the picture-dependent case
($C^{(s|t)}=0$ for $s,t=0,-1,-2$) such a RG flow cannot cancel the divergence
(39) in the partition function because the correlators
 (42) and (39) are differently normalized (the normalization of the latter
depends on N and therefore the flow (43) cannot remove the
divergence simultaneously for all the correlators).
  The resolution of this difficulty is that
the divergence (39) is removed not by the standard quadratic renormalization
of the $\lambda$-field  in the $(N+3)$-point correlator but by  the
cubic renormalization of the
space-time fields inside the $N+2$-point function
$\sim\lambda{A}H^n<V_5V_aV_{a_1b_1}...V_{a_Nb_N}>$, i.e. in the lower order
expansion term.
Indeed, using the transversality condition for the photon:
$k_aA^a(k)=0$ along with the momentum conservation, we can recast
the divergence (39) in the form:
\eqn\grav{\eqalign{
A_\Lambda(p_1,p_2;k_1,k_2;q_1...q_N)=
{i\over{N!}}
(p_2^a+k_2^a+\sum_{j=1}^N{q_j})
\lambda(p_1)\lambda(p_2)A_a(k_1)A_b(k_2)
\cr
{\prod_{j=1}^N}
H_{a_jb_j}(q_j)(1-\Xi^{-1}_\Gamma(N+2;4)F(N))S_{N+3}(k_1+p_1,p_2,k_2,q_1...q_N)
}}
Now consider the the $N+2$ order expansion term in the generating
functional (24) given by
\eqn\grav{\eqalign{A_{N+2}(p,k,q_1...q_N)={1\over{N!}}\int_{p,k,q}
{{\lambda(p)A_a(k)\prod_{j=1}^NH_{a_jb_j}(q_j)}}V_5(p)V^a(k)V^{a_jb_j}(q_j)}}
and the following operator-valued RG flows:
\eqn\grav{\eqalign{
\lambda(p_1)\rightarrow\lambda(p_1)+log\Lambda\sum_{n=-3;n\neq{-2,-1,0}}^\infty
{i}\alpha_n{p_1^a\lambda(p_1)}\int_{k,p}\lambda(p)A^a(k)
{\int_\Lambda}V_5^{(n)}(k+p)\cr
A_b(k_1)\rightarrow{A_b(k_1)}+log\Lambda\sum_{n=-3;n\neq{-2,-1,0}}^\infty
{i}\alpha_n{k_1^a{A_b(k_1)}}\int_{k,p}\lambda({p})A^a(k)
{\int_\Lambda}V_5^{(n)}(k+p)\cr
H_{a_jb_j}(q_j)\rightarrow{H_{a_jb_j}(q_j)}\cr
+log\Lambda
\sum_{n=-3;n\neq{-2,-1,0}}^\infty
{i}\alpha_n{q_j^a{H_{a_jb_j}(k_1)}}\int_{k,p}\lambda(p)
A^a(k)
{\int_\Lambda}
V_5^{(n)}(k+p)}}
where, as previously, $n$ labels the admissible pictures,
$\int_\Lambda$ is the worldsheet integral cut off at the $\Lambda$
scale, $\int_{k,p}$ stands for all the momentum integrals
 and $\alpha_n$ are some numbers yet to be determined.
The transformation of (45) under the RG flows (46) gives
\eqn\grav{\eqalign{A_{N+2}(k_1,p_1,q_1...q_N)\rightarrow
A_{N+2}(k_1,p_1,q_1...q_N)
\cr
+{i\over{N!}}
log\Lambda\int_{k_2,p_2}\sum_{n=-3;n\neq{-2,-1,0}}^\infty
\alpha_n(k_1^b+p_1^b+q_1^b+...q_N^b)\cr
\lambda(p_1)\lambda(p_2)
A_a(k_1)A_b(k_2)\prod_{j=1}^NH_{a_jb_j}(q_N)
\cr\times
<{|\rho_{(\Gamma,Z)}|^2}V_5^{(-3)}(p_1)V_5^{(n)}(k_2+p_2)V_a^{(-1)}(k_1)
V_b^{(-1)}(k_2)V_{a_1b_1}^{(-1)}(q_1)...V_{a_Nb_N}^{(-1)}(q_N)>}}
Now it is clear that the sum (47) will be truncated at
$n=N+3$ because the total ghost number of the correlator
(47) must be equal to $-2$ and all the terms in the  
$\rho_{(\Gamma,Z)}$-factor
have  positive ghost numbers; in the meantime all the correlators
with $n\leq{N+3}$ will contribute the factor
$\sim{S_{N+3}}(p_1,p_2+k_2,k_1,q_1...q_N)$
Thus the transformation of $A_{N+2}(k_1,p_1,q_1...q_N)$ under the flows
(46) can be expressed as
\eqn\grav{\eqalign{A_{N+2}(k_1,p_1,q_1...q_N)\rightarrow
A_{N+2}(k_1,p_1,q_1...q_N)\cr
+{i\over{N!}}
log\Lambda{\int_{k_2,p_2}}{S_{N+3}}(p_1,k_2+p_2,k_1,q_1..q_N)
\sum_{n=-3;n\neq{-2,-1,0}}^{N+3}
(k_1^b+p_1^b+q_1^b+...q_N^b)\cr
\lambda(p_1)\lambda(p_2)
A_a(k_1)A_b(k_2)\prod_{j=1}^NH_{a_jb_j}(q_N)
\sum_{n=-3;n\neq{-2,-1,0}}^{N+3}\alpha_n
}} 
Comparing this with (39) we conclude that
the RG flow of $A_{N+2}$ removes the divergence of (39) if we
require
\eqn\lowen{\alpha_{-3}+{\sum_{n=1}^{N+3}}\alpha_n
=1-\Xi_\Gamma^{-1}(N+4;2)F(N)}
leading to the following values of the $\alpha_n$:
\eqn\grav{\eqalign{\alpha_{n+3}
=\Xi_\Gamma^{-1}(n+4;2)F(n)-
\Xi_\Gamma^{-1}(n+5;2)F(n+1)\cr
n=0,1,2....\cr
\alpha_{-3}+\alpha_1+\alpha_2=1-\Xi_\Gamma^{-1}(4;2)F(0)
}}
The choice (50) of the coefficients in the RG flows (46)
 insures the unitarity of the
string perturbation theory and the absence of 
the UV divergences in the sigma-model (24).
Curiously, the series in $\alpha_n$ converges to
the inverse logarithm of the Feigenbaum's universality
constant $\delta=4.669...$ {\feigen}:
$$\alpha_{-3}+\sum_{n=1}^\infty\alpha_n{\approx}0.649\approx
{1\over{log\delta}}$$
Such a numerical coefficient (the inverse log of delta)
(which, roughly speaking, ``normalizes'' the $V_5$-operator
in the stochastic term)
 appears to be  universal
 in all the calculations involving the correlation
functions with picture-dependent vertices 
 In general, the $\beta$-functions in the
various  ghost-matter mixing 
backgrounds (corresponding to the brane insertions) always
contain the stochastic terms, universally normalized
regardless of the details of the ghost-matter mixing.
The underlying reasons for such a ``Feigenbaum universality''
in string theory are not clear at present; in principle, they
should be related to the chaotization of the RG flows
at certain critical values of $\lambda$ (which is related to the
viscosity and hence the Reynolds number of the ``stringy liquid''
- see the discussion below).
Some preliminary and incomplete attempts to explain this phenomenon have been 
made in {\ian}
This appears to be an interesting and 
intriguing question to clarify.

Finally, transforming to the position space
and adding the kinetic terms, we find
 the RG flows (46) to be given by
\eqn\grav{\eqalign{
{{\partial\lambda}\over{\partial\tau}}-({\vec{\nabla}})^2\lambda
+\lambda({\vec{A}}{\vec\nabla})\lambda\int_\Lambda{W_5(0)}=0\cr
{{\partial{\vec{A}}}\over{\partial\tau}}
-({\vec{\nabla}})^2{\vec{A}}
 +\lambda({\vec{A}}{\vec\nabla}){\vec{A}}\int_\Lambda{W_5(0)}=0\cr
{{\partial{H_{ab}}}\over{\partial\tau}}
-({\vec{\nabla}})^2{H_{ab}}
+\lambda({\vec{A}}{\vec\nabla}){H_{ab}}\int_\Lambda{W_5(0)}=0\cr
W_5(0)\equiv{\sum_{n=-3;n\neq{-2,-1,0}}^\infty{\alpha_nV_5^{(n)}(p=0)}
}
}}
where $\tau\equiv{log}\Lambda$.
In the case when the $\lambda$-field varies slowly,
the RG equations (51) become the equations of fluid mechanics
with viscosity $\nu\sim{1\over{\lambda}}$ and the velocity
\eqn\lowen{{\vec{v}}={\vec{A}}\int_\Lambda{W_5}}
The incompressibility of this liquid  follows trivially
from the transversality of the photon.
The second equation of (51) is the Navier-Stokes
equation for the velocity $\vec{v}$ operator.
It does not yet contain the term with the gradient of pressure.
In principle,the pressure term can be added as a result of the 
deterministic  contribution of the
OPE of two axions and gravitons the photon's $\beta$-function:
\eqn\lowen{
lim_{z,w\rightarrow{s}}V_{B,G}(k;z,\bar{z})V_{B,G}(p,w,\bar{w})
{\sim}|z-w|^{-2}(z+\bar{z}-w-\bar{w})^{-1}C^a(k,p)V_a(k+p)}
where
\eqn\lowen{C_a\sim{k}_a+...}
(here we skipped the terms cubic in the momentum)
This OPE particularly gives rise to the term $\sim{Tr(F(B+G)^2)}$
in the expansion of the DBI effective action
Then the pressure can be expressed in terms of the axion
and the graviton fields
\eqn\lowen{P\sim{Tr}(B+G)^2+...}
The random force term can also be obtained
if one considers the higher order expansion terms in $\lambda$
and $V_5$.
In particular, the term cubic in $\lambda$
contributes the stochastic 
term $\sim({\vec{\nabla}}\lambda)^2{\vec{\nabla}}\lambda
\int_\Lambda{V_5}$ to the photon's $\beta$-function and can be interpreted
as a random force acting in the liquid.
Finally, consider the last equation of the RG flows (51).
In case if the  $H$-field of (51) is a dilaton, i.e.
$H_{ab}=\varphi(\eta_{ab}-k_a{\bar{k}}_b-{\bar{k}}_ak_b)$,
this equation is simply the well-known equation for
the advection of the passive scalar by the 
turbulent liquid with the velocity $\vec{v}$ of (52) 
{\krai, \kraic, \gawed}
In case if $H$ is a 
graviton, the equation describes the evolution
of the stress-energy tensor of the fluid in the process of turbulence.
\centerline{\bf Discussion}
In this paper we have shown that
the renormalization group flows 
for the massless modes of NSR superstring in the presence
of D-branes have the form of the equations of fluid mechanics.
The velocity distribution in the turbulent flow was shown to be related to the 
photon (D-brane's U(1) field) multiplied by the D-brane creation
$W_5$-operator (52),(20).
The dilaton has been interpreted as a passive scalar 
(e.g. a temperature) while the roles of the axion and the graviton
in such a  string-turbulence glossary have been related to
the pressure and stress-energy tensor.
It should be stressed that such a string-to-turbulence
correspondence should be understood in terms of the off-shell
renormalization group flows, rather than the on-shell scattering
amplitudes. The results derived in this paper may suggest an isomorphism
between the off-shell superstring theory in the D-brane backgrounds
and the degrees of freedom of the fluid  mechanics, in the manner
of the well-known gauge-string correspondence {\ampf, \lpeq}.
Just like the confining QCD strings can be thought of being  made of the 
gluon fluxes confined to a thin tube, the 
strings of a fluid mechanics can be related (using
a bit simplistic language) to the
vortices of a turbulent liquid. 
Thus it may be tempting to use the language
and the formalism of the off-shell string theory to explore
the mechanism of turbulence (as the straightforward  analysis
of the Navier-Stokes equation is known to be extremely difficult).
In particular, one can hope to use the off-shell string theoretic
formalism to derive the Kolmogorov scaling or to study the
statistical  distribution
of the vortices.
In this paper we studied the evolution of the superstring massless
modes in the non-equilibrium curved backgrounds rather than
the stable ones created by static D-branes.These backgrounds appear when
, at the certain initial moment of time one introduces a D-brane into
the originally flat vacuum , by insertions of the $V_5$-operators
and the metric begins to fluctuate strongly.
Hydrodynamical character of the evolution of the massless modes 
is the important property of string dynamics in such a 
fluctuating metric.
Later on, however, the metric should be stabilized to the usual
D-brane like backgrounds, as, in the limit of infinite stochastic time
(corresponding to $\Lambda\rightarrow{0}$) the stochastic process 
(46), (51) reaches
the limit of thermodynamical equilibrium.
This means that the usual D-brane geometry (e.g. the
equilibrium configurations of the metric and the dilaton)
should be the result of certain soliton-type solutions of
the fluid mechanics equations. In particular,
if we believe that our 4-dimensional Universe lives 
on the asymptotically flat boundary of $AdS_5$
(which is suggested by the standard gauge-string correspondence)
one is led to a curious question whether the world we are living in has
been formed as a result of a certain giant multi-dimensional
turbulence in the early Universe. In other words, 
are we simply living on a hydrodynamical
soliton formed somewhere in the bulk of
a primordial fluid in higher dimensions? 

\centerline{\bf Acknowledgements}
It is a pleasure to thank Ali Chamseddine and 
Wafic Sabra for useful comments and discussions.

This paper is dedicated to the memory of Ian Kogan,
my collaborator and wonderful friend. Certain  initial ideas,
eventually evolved into some of the results of this work,
have been discussed almost a year ago
in our last phone talk with Ian,
 right
before his departure from Oxford to Trieste.
We had a productive conversation and agreed to continue
 the discussion few days later, after Ian's supposed
return.
Ian tragically passed away in Trieste
 on the 4th of June, 2003.

\listrefs
\end